\documentclass{osa-article}

\journal{ome}
\articletype{Research Article}
\usepackage[squaren,Gray]{SIunits}

\begin{document}

\title{Observation of topological valley hall edge states in honeycomb lattices of superconducting microwave resonators}

\author{Alexis Morvan,\authormark{1} Mathieu Féchant,\authormark{1} Gianluca Aiello,\authormark{1} Julien Gabelli,\authormark{1} and Jérôme Estève\authormark{1,*}}

\address{\authormark{1}Laboratoire de Physique des Solides, CNRS, Université Paris Saclay, Orsay, France}

\email{\authormark{*}jerome.esteve@u-psud.fr} %% email address is required

\begin{abstract}
We have designed honeycomb lattices for microwave photons with a frequency imbalance between the two sites in the unit cell. This imbalance is the equivalent of a mass term that breaks the lattice inversion symmetry. At the interface between two lattices with opposite imbalance, we observe topological valley edge states. By imaging the spatial dependence of the modes along the interface, we obtain their dispersion relation that we compare to the predictions of an \emph{ab initio} tight-binding model describing our microwave photonic lattices.   
\end{abstract}

\section{Introduction}

Even though static lattices for spinless particles that do not break time reversal symmetry have band structures that only contain bands with zero Chern number, topological effects can still be observed in such systems, with topological valley hall edge (TVHE) states being a prominent example \cite{xiao2007,semenoff2008,yao2008}. In a honeycomb lattice with an on-site energy imbalance $\mu$ between the nonequivalents $A$ and $B$ sites in the unit cell, the inversion symmetry is broken and a gap opens at the Dirac points, giving rise to an insulator \cite{Jackiw1976,semenoff1984}. By integrating the Berry curvature in the neighbourhood of each Dirac point, one obtains that each valley carries a topological charge $\pm1/2$, where the sign changes with the valley, the sign of $\mu$ and the band index \cite{xiao2007,yao2008}. If one considers two lattices with opposite imbalance $\mu$ connected along a boundary, the difference of topological charge for a given valley between the two sides of the boundary is one. The bulk-edge correspondence principle implies that two branches of edge states, one for each valley, must exist \cite{yao2009,delplace2011,mong2011}. The states are spatially localized along the boundary and their direction of propagation is correlated to the valley index, in a way similar to the quantum spin Hall effect, where the direction of propagation is correlated to the spin \cite{kane2005}. The two branches cross the gap and intersect in its center with a close to linear dispersion relation, whose slope is approximately given by the Fermi velocity \cite{semenoff2008}. Artificial photonic, phononic and acoustic lattices offer a perfect playground to design honeycomb lattices where TVHE states may be observed. The first experiments have been realized with sound waves \cite{lu2017,lu2018}, elastic waves \cite{vila2017}, microwaves \cite{wu2017,gao2018} and optical waves propagating in arrays of evanescently coupled waveguides \cite{noh2018}. 

Here, we report on the observation of TVHE states with microwave photons in lattices of superconducting resonators in the linear regime \cite{Underwood2012,underwood2016,owens:2018,dietz2018,kollar2019}. We have designed two samples with different boundaries, zigzag and armchair, between two lattices with an imbalance $\mu$, whose absolute value is equal to half the hopping amplitude between neighbouring sites. This results in the apparition of well localized TVHE states at the boundary that we image using a laser scanning technique \cite{morvan2020}. This allows us to reconstruct the dispersion relation of the states and to show that it is approximately linear, with a slope which is close to the Fermi velocity, independently of the type of boundary. We then compare our data to the more precise predictions of tight-binding models. The model parameters are obtained from an \emph{ab initio} model of the lattice and its predictions are in good agreement with our experimental data.            

\section{Experimental observation of topological valley hall edge states}
\subsection{Lattice design}
\begin{figure}
    \begin{center}
    \includegraphics{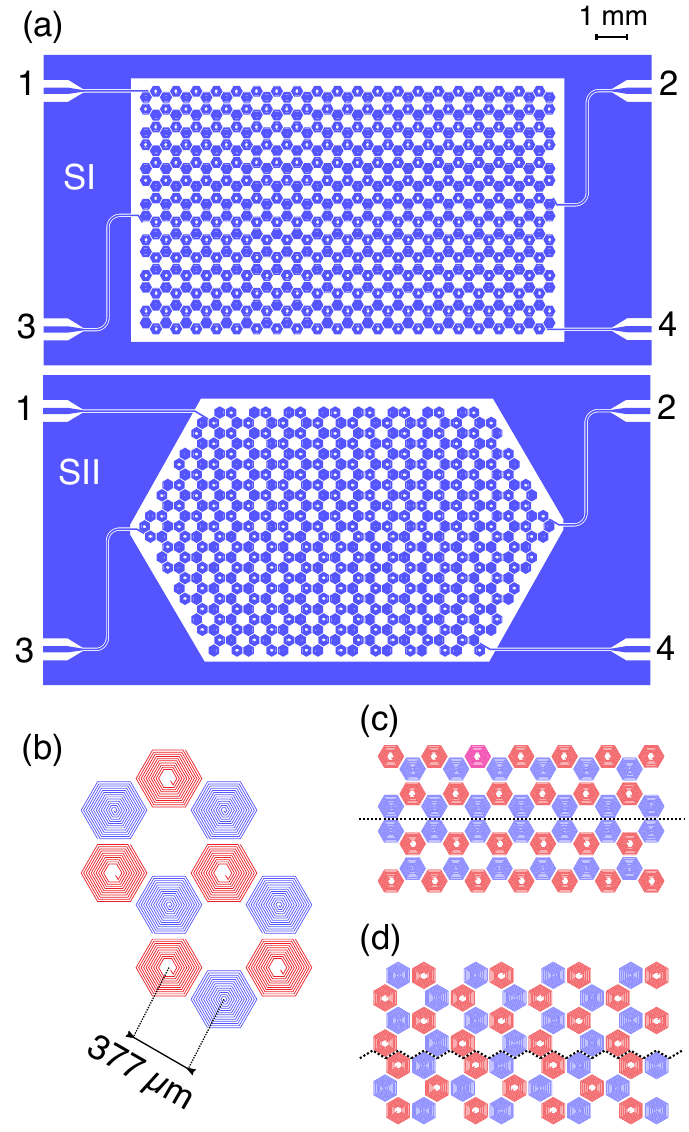}
    \caption{(a) Design of the two lattices studied in this article. Each sample is a honeycomb lattice where the $A$ and $B$ sites are occupied by spiral resonators with different lengths as shown in (b). The type of spiral occupying the $A$ or $B$ site is permuted between the lower and upper half of the sample at a horizontal boundary located close to the center of the sample. The SI and SII designs respectively correspond to a zigzag and an armchair boundary, as shown in (c) and (d). The SI design has 574 sites and the SII design has 480 sites. Four coplanar waveguides, labeled 1 to 4, couple the lattice to microwave ports that are connected to the measurement apparatus. Each waveguide is capacitively coupled to a single site located on the edge of the lattice. \label{fig:designs}}
    \end{center}
\end{figure}

The design of the two different samples, where we observe TVHE states are shown in figure \ref{fig:designs}. Locally, each sample is a honeycomb lattice, where each site is a spiral resonator. The site to site distance is $a=\unit{377}{\micro\meter}$. The nonequivalent $A$ and $B$ sites correspond to two spirals of slightly different length. The longer spiral has a fundamental resonance $\omega_0 \approx \unit{2\pi \times 6}{\giga\hertz}$, while the shorter one resonates at $\omega_0+ 2\mu$ with $\mu = \unit{2\pi \times 60}{\mega\hertz}$. The spirals are made of Nb on a Si wafer. More details about the resonator and fabrication techniques can be found in \cite{morvan2020}. This asymmetry breaks the lattice inversion symmetry and $\mu$ plays the role of a mass imbalance. Both samples are divided in two halves, with the lower and upper halves having an opposite imbalance. The sign of $\mu$ abruptly changes at a horizontal boundary in the center of the sample over one lattice site. The two designs SI and SII differ by the nature of the boundary: zigzag in sample SI and armchair in sample SII. Four coplanar waveguides are capacitively coupled to specific sites located on the sample edges and connect to outer ports that are used to probe the sample with a vector network analyzer. Ports 2\&3 are connected to sites located on both ends of the horizontal boundary, where TVHE states are expected, while ports 1\&4 are connected to corner sites. We also expect that edge states appear on the outer edges of the sample if they are zigzag or bearded terminated \cite{delplace2011}. In order to avoid hybridization between the TVHE states with these outer boundary edge states, the samples were designed to have no such states on the left and right edges, where the horizontal boundary begins and ends. This is the reason why the left and right edges of the SII sample are not vertical.

\subsection{Lattice transmission spectroscopy}
\begin{figure}
    \begin{center}
    \includegraphics{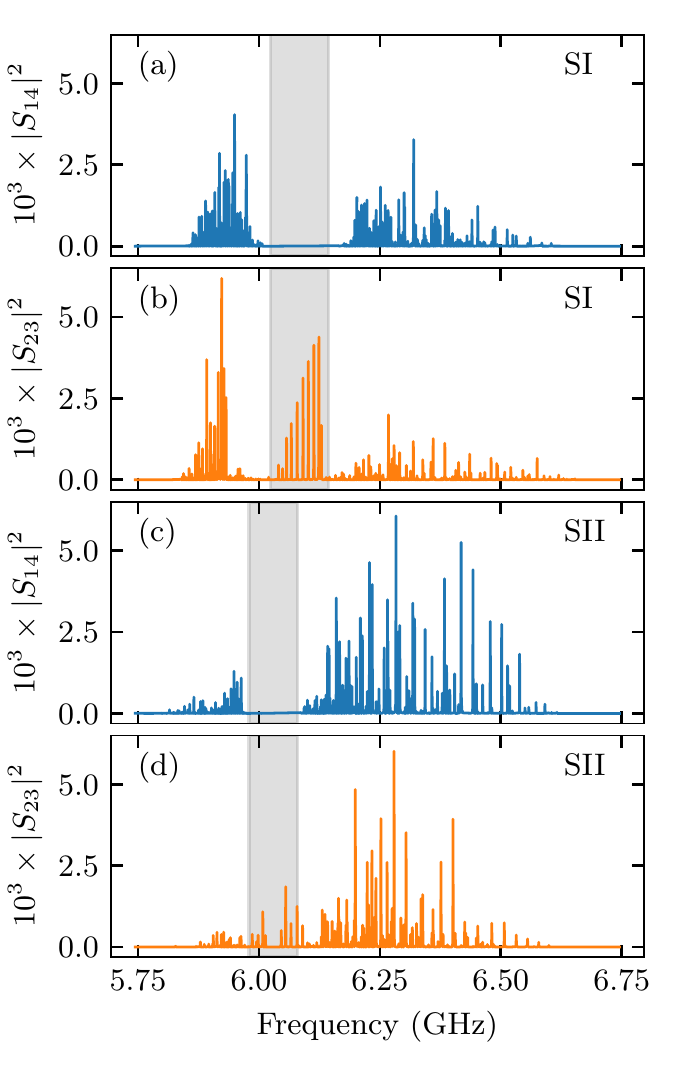}
    \caption{(a) Transmission $|S_{14}|^2$ through corner sites of sample SI. This measurement reveals the resonance frequencies of the lattice bulk modes that appear as sharp peaks. We observe two bands separated by a gap, the shaded area indicates the expected gap width $2\mu/(2\pi)=\unit{120}{\mega\hertz}$. (b) Transmission $|S_{23}|^2$ through two sites at the extremities of the boundary hosting TVHE states for the SI sample. The edge states appear as new resonant peaks not visible in (a) that lie in the gap. (c),(d) Same as (a) and (b) for the SII sample.
    \label{fig:transmission}}
    \end{center}
\end{figure}

Figure \ref{fig:transmission} shows different transmission spectra, for SI and SII samples, when the sample is cooled around \unit{1}{\kelvin}, way below the superconducting transition temperature of Nb. We observe a striking difference between the transmission spectra from one site to an other, when the two sites located at the corner of the sample (fig. \ref{fig:transmission}a,c) or at the horizontal boundary (fig. \ref{fig:transmission}b,d). In figures \ref{fig:transmission}a and \ref{fig:transmission}c, which respectively corresponds to sample SI and SII, the resonance peaks correspond to bulk modes of the lattice: we clearly identify two bands separated by a gap with a width close to the expected value 2$\mu$. The transmission maxima are much smaller than one, indicating that the modes are under-coupled: intrinsic losses dominate the coupling losses to the measurement waveguides. In figures \ref{fig:transmission}b and \ref{fig:transmission}d, we observe new resonances in the gap that we attribute to TVHE states. Some of these peaks have a higher transmission maximum in comparison to average bulk modes. This is a first hint that these states are localized at the boundary and have a large weight on the two sites connected to ports 3\&4. Boundary states are expected to couple more to these ports than bulk modes, which results in a higher transmission at resonance because the modes are under-coupled.  

\subsection{Edge state mode imaging}
\begin{figure}
    \begin{center}
    \includegraphics{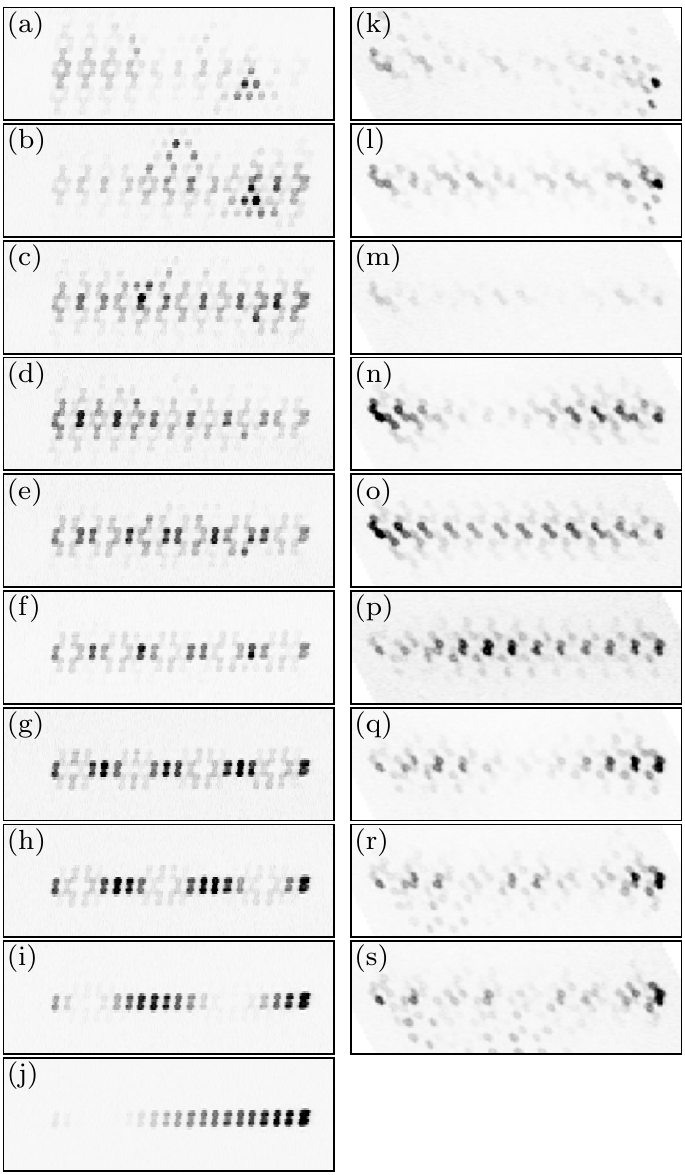}
    \caption{Intensity maps of TVHE states for sample SI on the left (a-j) and sample SII on the right (k-s). The intensity of a mode on a given pixel corresponds to the measured transmission drop at the mode frequency, which is induced by a focused laser spot that is scanned across the sample. Here, we only show the resulting images in a horizontal band centered around the boundary. The frequency of the modes increases from $\unit{6.04}{\giga\hertz}$ to $\unit{6.13}{\giga\hertz}$ for images (a) to (j), and from $\unit{5.97}{\giga\hertz}$ to $\unit{6.08}{\giga\hertz}$ for images (k) to (s).
    \label{fig:images}}
    \end{center}
\end{figure}

In order to confirm that the modes appearing in the gap are TVHE states, we use a laser scanning technique to measure the spatial dependence of these modes across the lattice \cite{underwood2016,wangH2019}. The image of one mode is obtained by recording the transmission loss at the frequency of the mode induced by a laser spot focused onto the sample as a function of the position of the spot. The laser beam direction is steered by a motorized mount located outside the cryostat. The beam is relayed by two lenses through the thermal shields and is focused onto the sample by a final lens. The laser induces losses that are proportional to the local current density in the sample and thus to the local mode intensity (see \cite{morvan2020} for details). Figure \ref{fig:images} shows the intensity for the modes that we identify as TVHE states, because they only appear in $|S_{23}|^2$, are absent from the bulk transmission $|S_{14}|^2$ and resonate inside the gap. The images show that these modes are indeed localized at the boundary. Along the longitudinal direction parallel to the boundary, the mode profiles have a periodic behaviour. This standing wave pattern comes from the interference of TVHE modes with the same energy and opposite wavevectors. These states are in different valleys, but because the coupling to the excitation waveguide is not valley selective, modes propagating in both directions along the boundary are excited.

\begin{figure}
    \begin{center}
    \includegraphics{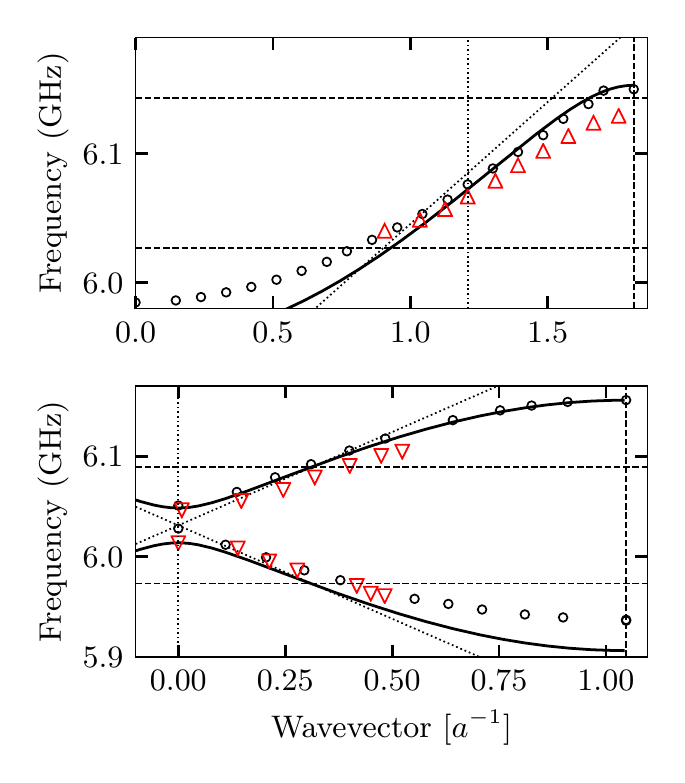}
    \caption{Dispersion relation of the TVHE states observed in the SI (top) and SII (bottom) sample. The wavevectors are obtained from a Fourier analysis of the images shown in figure \ref{fig:images}. Data points are shown as red triangles, an upwards triangle means that the measured wavevector has been unfolded in the Brillouin zone (see text). The vertical dashed line indicates the edge of the boundary Brillouin zone. The vertical dotted line indicates the position of the lattice Dirac point and the horizontal dotted lines the gap. The tilted dotted line corresponds to a linear dispersion around the Dirac points with a velocity $v_F$. The black solid line is the analytical prediction from a NN tight-binding model, while the black circles correspond to a numerical simulation of the TVHE states including NNN coupling and finite size effects. 
    \label{fig:dispersion}}
    \end{center}
\end{figure}

An important prediction for the TVHE states is that their dispersion relation in the middle of the gap is close to linear, with a velocity approximately equal to the Fermi velocity. In order to experimentally test this prediction, we Fourier transform the longitudinal profiles obtained from the images shown in figures \ref{fig:images}. More precisely, we compute one spectrum for the average intensity over the two lines of sites immediately below and above the boundary and a second spectrum for the average intensity over the two lines which are one site away from the boundary (one below and one above). We then average the square modulus of these two spectra and obtain one spectrum per image. Because we measure the intensity of the modes and not the amplitude, a superposition of modes with wavevectors $k$ and $-k$ results in peaks at $2k$ and $-2k$, which are then eventually folded back in the first Brillouin zone if $2k>\pi/a_b$, where $a_b$ is the period of the boundary. For example, an image with a low spatial frequency, as shown in figure \ref{fig:images}(j), may correspond to modes with $\pm k \approx 0$ or $\pm k \approx \pm \pi/a_b$. In order to lift this ambiguity, we have to suppose that the observed modes have a spatial dependence close to the expected ones. For the zigzag boundary, we expect that most peaks have a high wavevector and therefore we unfold all the measured values. For the armchair boundary, we expect the opposite and do not unfold any value. The results are shown as triangles in figure \ref{fig:dispersion}. 

\begin{figure}
    \begin{center}
    \includegraphics{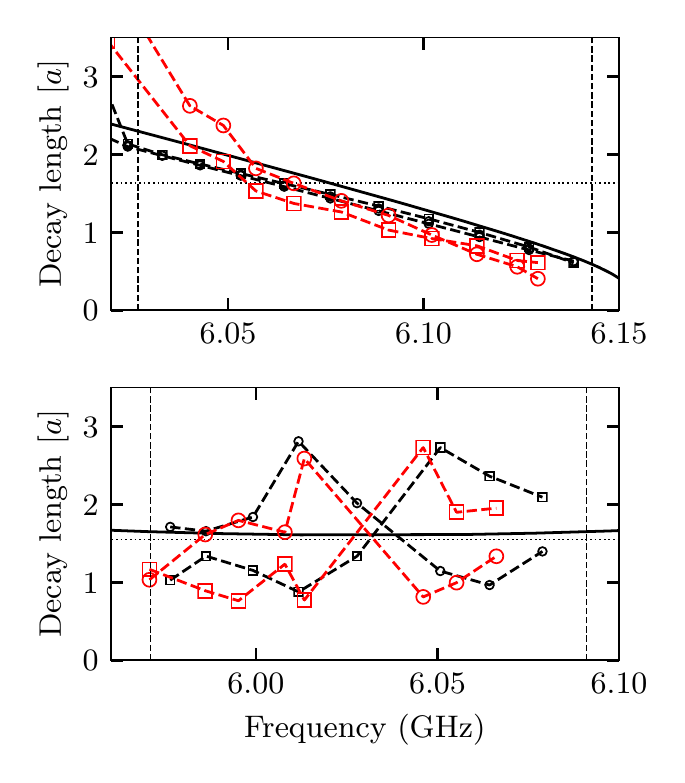}
    \caption{Decay length of the TVHE states observed in the SI (top) and SII (bottom) sample as a function of their frequency. The data points are shown in red, the two different markers identify the decay length for each sublattice. The vertical dashed lines identify the gap. The dotted line is the prediction $\xi/a=3 t_1/(4 \mu)$ of the Dirac equation. The black solid line is the analytical prediction from a NN tight-binding model, while the black symbols correspond to a numerical simulation of the TVHE states of the two samples, including NNN coupling and finite size effects. As for the data, the two different symbols correspond to the two different sublattices. 
    \label{fig:decay}}
    \end{center}
\end{figure}

Finally, we compute a transverse profile for each mode in order to evaluate the decay length of the TVHE states with the distance to the boundary. For each image, we obtain two transverse profiles by averaging over the sites occupied by one or the other spiral. We then fit both profiles to an exponential decay and obtain two decay lengths, one for each sublattice. The data are shown with red markers in figure \ref{fig:decay}. In the zigzag case, we observe that the two decay lengths are almost equal, but this is not the case for the armchair boundary. The same feature is observed in the simulation of the sample, which we detail in the next section.

\section{Comparison with massive Dirac equation and tight-binding model predictions}
In \cite{semenoff2008}, Semenoff \emph{et al.} derive the characteristics of the boundary states by using a massive Dirac equation to describe the propagation in the lattice. Supposing that the states have a wavector $q$ in the longitudinal direction $x$ along the boundary, the dispersion relation $\varepsilon(q)$ and the transverse wavefunction $\varphi(y)$ is obtained by solving the eigenvalue problem
\begin{equation}
    \left( -i\tau_z \otimes \sigma_x \, v_F \frac{\partial}{\partial y} + v_F q \, \tau_Z \otimes \sigma_y  \pm \mu \,\tau_z \otimes \sigma_z \right) \varphi(y) = \varepsilon \, \varphi(y)
\end{equation}
where the $\sigma_i$ and $\tau_i$ are the usual Pauli matrices used to describe Dirac matter \cite{cayssol2013} and the sign of $\mu$ changes at the boundary located at $y=0$. One obtains two solutions 
\begin{equation}
    \varphi_+(y) = e^{- \mu |y| / v_F  } \begin{bmatrix} 1 \\ i \\ 0 \\ 0 \end{bmatrix} \ \ \ \ \ \varphi_-(y) = e^{- \mu |y| / v_F  } \begin{bmatrix} 0 \\ 0 \\ 1 \\ -i \end{bmatrix}
\end{equation}
with corresponding eigenvalues $\varepsilon_{\pm}= \pm v_F q$. From these expressions, we obtain a linear dispersion of the TVHE states with a velocity equal to $v_F$ and a constant localization length $\xi = v_F/(2 \mu)$. We also note that the states have equal weights on the two sublattices. 

These simple predictions are plotted as dotted lines in figures 4\&5. In order to obtain values for $\mu$ and $v_F$, we use the tight-binding model that we developed in \cite{morvan2020}. The parameters of the model are computed from a coupled mode theory approach \cite{elnaggar2015}: we look for a solution for the electric and magnetic fields across the lattice as a linear combination of the fundamental mode of the spirals. The coupling between the modes is given by the overlaps between the electric and magnetic fields of the modes. In our lattice, the electric and magnetic couplings have similar strength and add up to contribute to the coupling between neighboring sites. The resulting eigenvalue problem can be transformed to a tight-binding form as usually done for electrons in solid. 
We obtain a nearest-neighbour (NN) coupling $t=2\pi \times \unit{125}{\mega\hertz}$, next nearest-neighbour (NNN) coupling $t_2=2\pi \times \unit{20}{\mega\hertz}$ and an imbalance $\mu = 2 \pi \times \unit{60}{\mega\hertz}$. Neglecting the NNN coupling, the analogue of the Fermi velocity is then given by $v_F = 3 t a/2$, which corresponds to a group velocity of approximately $\unit{10^6}{\metre\per\second}$. The decay length in unit of the lattice spacing is equal to $\xi/a = 3 t/(4 \mu) \approx 1.6$.

\begin{figure}
    \begin{center}
    \includegraphics{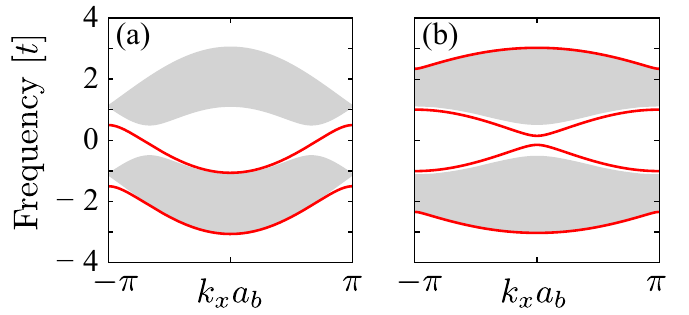}
    \caption{Tight binding model predictions for the dispersion of TVHE states (red lines) for the zigzag boundary (a) and the armchair boundary (b) as a function of the wavevector $k_x$ along the boundary. The period of the boundary $a_b$ is $\sqrt{3} a$ for the zigzag boundary and $3a$ for the armchair one. The NN coupling is such that $t=2\mu$, NNN coupling is neglected. The grey shaded area indicates the two bulk bands.
    The red lines for the edge states inside the gap correspond to the solid black lines shown in figure \ref{fig:dispersion}.
    \label{fig:band}}
    \end{center}
\end{figure}

The Dirac equation approach is valid when $\xi/a$ is large, which is not the case here. However, we find that its predictions agree reasonably well with the data. The main discrepancies are a slight overestimate of the group velocity for the zigzag edge states and the prediction of a constant decay length, which is not observed in both samples. More refined predictions can be obtained by considering a NN tight-binding description of the boundary problem and looking for periodic solutions along the boundary, which exponentially decay away from the boundary, as can be done for edge states in graphene \cite{wakabayashi2010,sasaki2011}. Calculations for both the zigzag and the armchair boundary are detailed in the supplementary material and the results are shown in figure \ref{fig:band}. The calculated band structure only depends on the ratio $t/\mu$, which is here equal to 2. 

In the case of the zigzag boundary, two sets of edge states are obtained with only one that crosses the gap and corresponds to the TVHE states predicted by the Dirac equation. Simple analytical expressions are obtained for the dispersion relation and for the decay length as derived in the supplemental document. The frequency of the TVHE states as a function of the wavevector $k_x$ along the boundary is given by 
\begin{equation}
    \omega(k_x) = t -\sqrt{\mu^2 + 4 t^2 \cos^2 \sqrt{3} k_x a/2 } \label{eq:zigzag1}
\end{equation}
Expanding this dispersion relation around  $k_x = -2\pi/(3 \sqrt{3}a) + q$  and $k_x = 2\pi/(3 \sqrt{3}a) + q$, two branches of states with linear dispersion are obtained
\begin{equation}
    \omega_\pm(q) = t -\sqrt{\mu^2 + t^2} \pm  \frac{3 t^2}{2\sqrt{\mu^2+t^2}} q a  \label{eq:zigzag2}
\end{equation}
These branches coincide with the predictions of the Dirac equation model when $\mu/t$ is small ($\xi/a$ large), in which case $\omega_\pm(q) \approx \pm v_F q$. One also sees that the velocity is reduced below $v_F$ taking into account corrections of order $(\mu/t)^2$. The prediction for the localization length is
\begin{equation}
    \xi(q) = -\frac{3 a}{4 \ln \lambda} \ \ {\rm with} \ \ \lambda \approx -1 + \frac{\mu}{t} \left( 1 + \frac{3qa}{2} \right)
\end{equation}
where we already made the $\mu/t$ expansion and the expansion around the Dirac point. At $q=0$, we recover $\xi(0)/a = 3 t /(4 \mu)$. Away from the Dirac point, we find that $\lambda$ increases linearly with $q$, leading to a decay length that linearly decreases with $q$ as observed in figure \ref{fig:decay}. These predictions (without any expansion) correspond to the solid black lines in the top plots of figures \ref{fig:dispersion}\&\ref{fig:decay}. 

In the case of the armchair boundary, four sets of states are obtained with two sets that correspond to the expected TVHE states. The calculation of the dispersion and localization lengths is more complicated and requires to numerically solve equations that can be found in the supplemental document. The results obtained for the two branches crossing the gap correspond to the solid black lines in the bottom plots of figures \ref{fig:dispersion}\&\ref{fig:decay}. A specific feature of the armchair case is that a small gap appears when the two branches of edge states cross at $k_x=0$ as can be seen in figure \ref{fig:dispersion}\&\ref{fig:band}. As argued in \cite{noh2018}, this difference between the zigzag and armchair TVHE states comes from the fact that the boundary mixes the valley in the armchair case but not in the zigzag case. This valley mixing couples counter-propagating states near $k_x=0$ and a gap opens. This gap is of order $\mu^2/t$ and tends to zero when $\mu/t$ becomes large and the Dirac equation predictions are recovered. In our case, the gap, which is visible in the simulation, is on the order of the level spacing between the different states. Therefore, except for the two states with zero wavevector, its presence is barely visible, and most of the states follow an close to linear dispersion relation. 

Finally, we have performed a full numerical simulation of the two samples including NNN coupling and the exact sample geometry in order to take into account finite size effects. We identify and analyze the TVHE states following the same procedure as for the experimental data and obtain the black points shown in figures \ref{fig:dispersion}\&\ref{fig:decay}. These simulations confirm the pertinence of the analytical calculation for an infinite lattice neglecting NNN coupling. In the case of the armchair boundary, the simulations partly reproduce the variation of the decay length with frequency, which we therefore attribute to finite size effects. These effects are more pronounced than for the zigzag sample because of the cropped regions on the left and right sides of the sample.  

\section{Conclusion}
In conclusion, we have observed topological valley hall states with microwave photons in a lattice of superconducting resonators. This work validates that lattices with well tailored properties can be designed with this technology and that simple tight-binding models accurately describe their properties. The parameters entering the model are obtained from \emph{ab initio} numerical simulations of the electromagnetic field of a single resonator. In a future work, it could be interesting to realize a valley selective excitation in order to observe the valley Hall effect where the direction of propagation along the boundary could be selected via the valley index. This requires addressing at least two sites on the two sublattices close to the boundary, which is challenging with a single planar circuit as considered here but could be done with the recent multi-layer approach developed in circuit QED. Finally, lattices of superconducting resonators offer the possibility to introduce a controlled non-linearity, which can be as large as the hopping amplitude, with the perspective to study the interplay between topological and non-linear effects  \cite{schmidt2013,fitzpatrick2017,bleu_robust_2018,ma2019,smirnova2020}. 

\section*{Acknowledgments}
We thank Gilles Montambaux and Marco Aprili for fruitful discussions. We thank Jean-Noël Fuchs for giving us the idea to perform these experiments and for fruitful discussions.  

\section*{Disclosures}
The authors declare no conflicts of interest.

\medskip

\noindent See Supplement 1 for supporting content.

\bibliography{bibliography}

\end{document}